\documentclass[10pt,conference]{IEEEtran}
\IEEEoverridecommandlockouts

\newif\ifDEBUG
\DEBUGfalse

\usepackage{cite}
\usepackage{float}
\usepackage{graphicx}
\usepackage{textcomp}
\usepackage[dvipsnames]{xcolor}

\usepackage{makecell}
\usepackage[longtable]{multirow}
\usepackage{colortbl}
\usepackage{longtable}
\usepackage{booktabs}

\def\BibTeX{{\rm B\kern-.05em{\sc i\kern-.025em b}\kern-.08em
    T\kern-.1667em\lower.7ex\hbox{E}\kern-.125emX}}
\usepackage{booktabs} 
\usepackage{amsmath}
\usepackage{algorithm}
\usepackage{algpseudocode}
\usepackage[normalem]{ulem} 
\usepackage{xspace}
\usepackage{multirow} 
\usepackage{balance} 
\usepackage{fancyvrb} 
\usepackage{caption}

\usepackage{enumitem}
\setlist[itemize]{leftmargin=*,noitemsep,topsep=0pt}
\setlist[enumerate]{leftmargin=*}

\usepackage{etoolbox}
\makeatletter
\patchcmd{\@makecaption}
	{\scshape}
	{}
	{}
	{}
\makeatletter
\patchcmd{\@makecaption}
	{\\}
	{.\ }
	{}
	{}
\makeatother
\def\tablename{Table}

\newcommand{\code}[1]{\texttt{{\small #1}}}

\newcommand{\eg}{\textit{e.g.,}\xspace}

\usepackage{mdframed}
\mdfsetup{skipabove=0.5\topskip,skipbelow=0.5\topskip,align=center}

\usepackage{amsthm}
\newtheorem{thm}{Theorem}

\newcommand{\benchname}[1]{\emph{#1}}

\DeclareMathSymbol{\mlq}{\mathord}{operators}{``}
\DeclareMathSymbol{\mrq}{\mathord}{operators}{`'}

\newif\ifSAVESPACE
\SAVESPACEfalse

\ifSAVESPACE

\else

\fi

\usepackage{todonotes}
\usepackage{soul}
\usepackage{fontawesome}
\usepackage{ragged2e}

\ifDEBUG
    \newcommand{\AH}[1]{\todo[color=cyan,inline]{AH:#1}}
    \newcommand{\AM}[1]{\todo[color=red,inline]{Machiry:#1}}
    \newcommand{\JD}[1]{\todo[color=yellow,inline]{JD:#1}}
    \newcommand{\SA}[1]{\todo[color=green,inline]{SA:#1}}
    \newcommand{\PA}[1]{\todo[color=orange,inline]{PA:#1}}
    
    \newcommand{\KR}[1]{\todo[color=yellow,inline]{Kyle:#1}}
    \newcommand{\LS}[1]{\todo[color=green,inline]{LS:#1}}
    \newcommand{\HP}[1]{\todo[color=green,inline]{HP:#1}}
    \newcommand{\NJE}[1]{\todo[color=red,inline]{NJE: #1}}
    \newcommand{\GKT}[1]{\todo[color=red,inline]{GKT:#1}}
    \newcommand{\KL}[1]{\todo[color=cyan,inline]{KL:#1}}
    \newcommand{\RH}[1]{\todo[color=red,inline]{RH:#1}}
    \newcommand{\WJ}[1]{\todo[color=SkyBlue,inline]{Wenxin:#1}}    
    
\else
    \newcommand{\AH}[1]{}
    \newcommand{\AM}[1]{}
    \newcommand{\JD}[1]{}
    \newcommand{\SA}[1]{}
    \newcommand{\PA}[1]{}
    \newcommand{\KR}[1]{}
    \newcommand{\LS}[1]{}
    \newcommand{\HP}[1]{}
    \newcommand{\NJE}[1]{}
    \newcommand{\GKT}[1]{}
    \newcommand{\KL}[1]{}
    \newcommand{\RH}[1]{}
    \newcommand{\WJ}[1]{}
    
\fi

\usepackage{hyperref}

\usepackage{cleveref}
\crefformat{section}{\S#2#1#3}
\crefname{figure}{Figure}{Figures}
\crefname{table}{Table}{Tables}
\crefname{theorem}{Theorem}{Theorems}
\crefname{thm}{Theorem}{Theorems}
\crefname{lemma}{Lemma}{Lemmata}
\crefname{equation}{Eqt.}{Eqts.}
\crefformat{Grammar}{Grammar #1}
\crefname{appendix}{Appendix}{Appendices}
\crefname{listing}{Listing}{Listings}

\usepackage{url}

\usepackage{tcolorbox}

\makeatletter
\newcommand{\linebreakand}{%
  \end{@IEEEauthorhalign}
  \hfill\mbox{}\par
  \mbox{}\hfill\begin{@IEEEauthorhalign}
}
\makeatother

\usepackage{pifont}

\newcommand{\bul}[1]{\textbf{\underline{#1}:}}

 \usepackage{pifont}

\newcommand{\better}[1]{\textcolor{ForestGreen}{#1}\xspace}
\newcommand{\worse}[1]{\textcolor{Maroon}{#1}\xspace}

\begin{document}

\title{Towards Energy-Efficient Code Optimization With Large Language Models}
\title{Large Language Models for Energy-Efficient Code: Emerging Results and Future Directions}


\author{
\IEEEauthorblockN{Huiyun Peng, Arjun Gupte, \\Nicholas John Eliopoulos, Chien Chou Ho,\\ Rishi Mantri, Leo Deng*,\\ Wenxin Jiang, Yung-Hsiang Lu}
\IEEEauthorblockA{
\textit{Department of Electrical}\\ 
\textit{and Computer Engineering} \\
\textit{Department of Computer Science*}\\
\textit{Purdue University} \\
West Lafayette, Indiana, USA
} \\

\and
\IEEEauthorblockN{Konstantin Läufer,\\ George K. Thiruvathukal}
\IEEEauthorblockA{
\textit{Department of Computer Science} \\
\textit{Loyola University Chicago}\\
Chicago, Illinois, USA \\
}
\and
\IEEEauthorblockN{James C. Davis}
\IEEEauthorblockA{
\textit{Department of Electrical}\\ 
\textit{and Computer Engineering} \\ 
\textit{Purdue University} \\ West Lafayette, Indiana, USA \\
}
}
\IEEEaftertitletext{\vspace{-3\baselineskip}} 

\maketitle

\begin{abstract}
Energy-efficient software helps improve mobile device experiences and reduce the carbon footprint of data centers.
However, energy goals are often de-prioritized in order to meet other requirements.
We take inspiration from recent work exploring the use of large language models (LLMs) for different software engineering activities.
We propose a novel application of LLMs: as code optimizers for energy efficiency.
We describe and evaluate a prototype,
  finding that over 6 small programs our system can improve energy efficiency in 3 of them, up to 2x better than compiler optimizations alone.
From our experience, we identify some of the challenges of energy-efficient LLM code optimization and propose a research agenda. 

\end{abstract}


\section{Introduction}
Energy efficiency has become a critical issue across various domains, from mobile devices to large-scale data centers.
\WJ{Just thinking, is there another example we can put here showing that energy efficiency can have worse impacts than just shorter battery life? Maybe cyber-physical system? Possible reference: \url{https://ieeexplore.ieee.org/stamp/stamp.jsp?arnumber=6006498&casa_token=CqwQ6fMh8OgAAAAA:WeEcSa9_c6IJx1ColYHSULxc4njHKgEBKw4_k_PyQR5tjm_akCCMAVurMbMfxMeL1YzonPIm4ko&tag=1}
}
On mobile devices such as those running Android, energy efficiency directly impacts battery life and accessibility, making it a key concern for user experience\cite{Wilke2013}.
On a larger scale, data centers contribute significantly to climate change, accounting for 4\% of electricity generation in the United States\cite{powering_intelligence2024} and 3\% in the European Union\cite{eu2024sustainability}.
Improving computing energy efficiency is part of addressing environmental sustainability.
Previous approaches have focused primarily on physical hardware, with limited discussion on improving the underlying software for energy efficiency~\cite{Manotas2016}. Efforts to design energy-efficient programs have introduced energy models~\cite{Aroca2014, Gao2014, Dayarathna2015, Daradkeh2022}, energy measurement tools~\cite{Almeida_Arteaga_Blanco_Cabrera_2015}, and energy-aware design patterns~\cite{Sahin2012, Noureddine2015, Pinto2016, Maleki2017}.
However, the barrier to adopting these energy-efficient practices remains high, often introducing complexity accessible only to systems experts~\cite{Papadopoulos2018}.
Software engineers recognize energy efficiency as a desirable property~\cite{Karita2019}. However, it often loses out to organizational goals like latency and throughput~\cite{Manotas2016}, due to the ineffectiveness of existing methods in meeting the dynamic nonfunctional performance requirements.
LLMs like ChatGPT are transforming software engineering by aiding in tasks like debugging and code optimization~\cite{Ozkaya2023, sarkar2022likeprogramartificialintelligence}. 
Although their potential for energy efficiency has yet to be fully explored, LLMs show promise as powerful aids in optimizing code for energy-aware practices.

\begin{figure}[t]
    \centering
    \includegraphics[width=0.9\columnwidth]{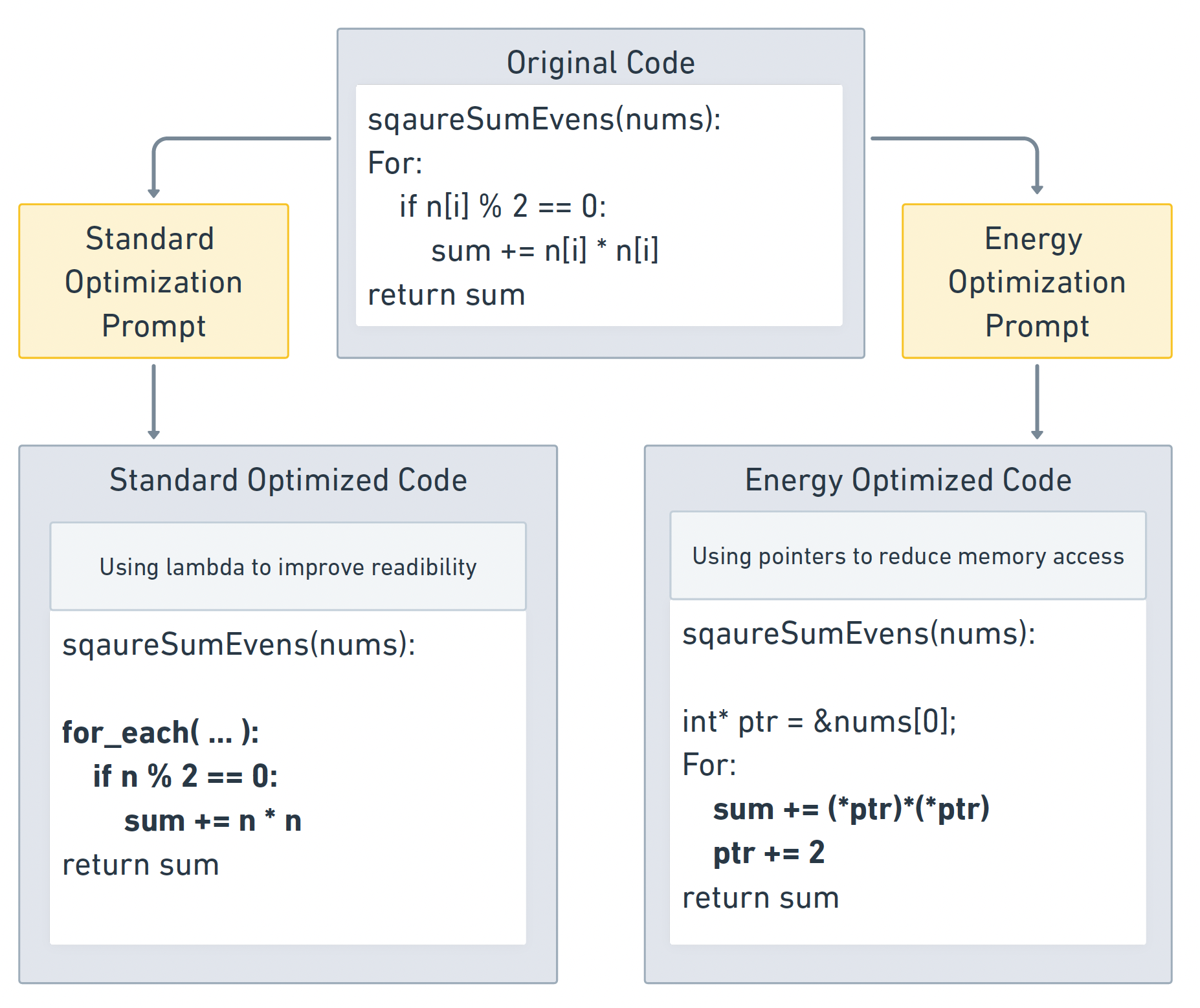}
    \caption{
    Standard optimization prompts focus on speed, memory optimization, and code readability. 
    Our proposed energy optimization prompts focus on reducing energy usage. 
    One must then ensure that optimizations \textit{actually} reduce energy use, not just performance metrics such as latency and FLOPS.
    }
    \label{fig:example_code_snippet}
\end{figure}

In this paper, we explore a novel approach to energy-efficient software development.
We ask:
\textit{Can LLMs assist developers in optimizing energy efficiency without compromising performance or correctness?}
As shown in \cref{fig:example_code_snippet}, we propose an automated tool designed to refactor software with a primary focus on optimizing energy efficiency, going beyond conventional performance metrics to directly target energy usage reduction.
This tool incorporates energy-aware prompts that are input into a generator LLM alongside the original code to produce more efficient program outputs.
The optimized code is then evaluated and refined through Natural Language Feedback from an evaluator LLM, allowing for iterative improvements to the code itself.
This approach offers the advantage of being portable --- it is easy to set up, and compatible with multiple programming languages without significant modifications.
We present preliminary results from evaluating our prototype using the \JD{is this the right benchmark name? I thought it was computer lang bench or somethign} \textit{Energy-Language} benchmark in \cref{sec:methodology}.
In \cref{sec:plan} we outline a research agenda for extending these findings.


\section{Background and Related Works} \label{sec:background}
\subsection{Energy-efficient Computing and Software}
\label{subsec:energy efficiency}
Energy-efficient computing focuses on reducing power consumption, usually while maintaining performance or correctness.
Current solutions for energy-efficient software frequently require heavyweight design approaches~\cite{Brinke2013, Brinke2014}, pattern catalogues~\cite{Pinto2016, Maleki2017}, specialized programming languages~\cite{Couto2017}, or decision frameworks~\cite{Manotas2014}, which makes it difficult for software engineers to adopt, implement, and subsequently maintain energy characteristics after evolution.
\NJE{citation?}

Another challenge of energy-efficient software design is that relationships between memory, latency, and energy are counterintuitive.
Common metrics for code performance such as FLOPS, latency, and memory usage are not necessarily strongly correlated 
with each other or with energy efficiency~\cite{efficiency_misnomer_2022}.
On certain hardware, even reducing input sizes may increase latency~\cite{prune_one_more_2024}.
Furthermore, improving characteristics such as parallelism~\cite{Mondal2015} may increase energy usage~\cite{Jin2017}.
\textit{Thus, creating code that avoids confounding performance metrics with energy efficiency is difficult, but critically important if energy use is the primary metric of interest.}

\subsection{LLM-Driven Code Generation and Optimization}
LLMs are transforming software engineering practices~\cite{Ozkaya2023, sarkar2022likeprogramartificialintelligence}.
Recent experiments show LLMs assisting with error message interpretation~\cite{Leinonen_2023}, cybersecurity defect repair~\cite{Pearce2023}, cloud incident mitigation~\cite{Ahmed2023}, and requirements elicitation~\cite{white2023chatgptpromptpatternsimproving}.
Moreover, recent studies also evaluate LLMs on efficient code generation and optimization, showing they can achieve both without compromising correctness~\cite{liu2024evaluatinglanguagemodelsefficient, shypula2024learning, Fan2023}. 
However, existing research primarily focuses on leveraging LLMs to generate fast code, rather than energy-efficient code. When energy consumption is the primary concern, current code-generation LLMs may fall short, as they do not explicitly consider the impact of their code on power consumption. \textit{Nevertheless, building on previous work, we anticipate that with appropriate adaptations, LLMs can be made to optimize for energy efficiency, extending their capabilities beyond just minimizing latency.}
\section{Prototype: Design, Implementation, \& Eval.} \label{sec:methodology}
\WJ{Should we be consistent with ``optimization'' and ``refactoring''?}
We propose an approach to evaluate LLMs' potential in refactoring software for energy efficiency. As an initial test of this concept, we developed a prototype of an automated LLM-assisted tool for energy optimization. 
To ensure effectiveness, all optimizations must maintain the original system behavior without introducing semantic changes while improving energy efficiency.
This section outlines the design, prototype implementation, and preliminary evaluation of our approach.

\subsection{Design}
Our system leverages state-of-the-art prompt engineering techniques and LLM feedback loops, applying them uniquely to the domain of energy measurement.
The key to our design lies in enabling the LLM to be energy-aware, guiding it to generate code that not only optimizes for speed but also makes significant improvements in energy efficiency.
As we described in \cref{subsec:energy efficiency}, optimizing a program for energy efficiency is non-trivial, since metrics associated with latency and throughput (parallelism) are not necessarily correlated with power consumption.
To overcome this challenge, we incorporate energy profiling mechanisms into the LLM feedback loop.
The LLM is given power-consumption data alongside latency, which is used to adjust generated-code.

\begin{figure}[t!]
    \includegraphics[width=0.9\columnwidth]
    {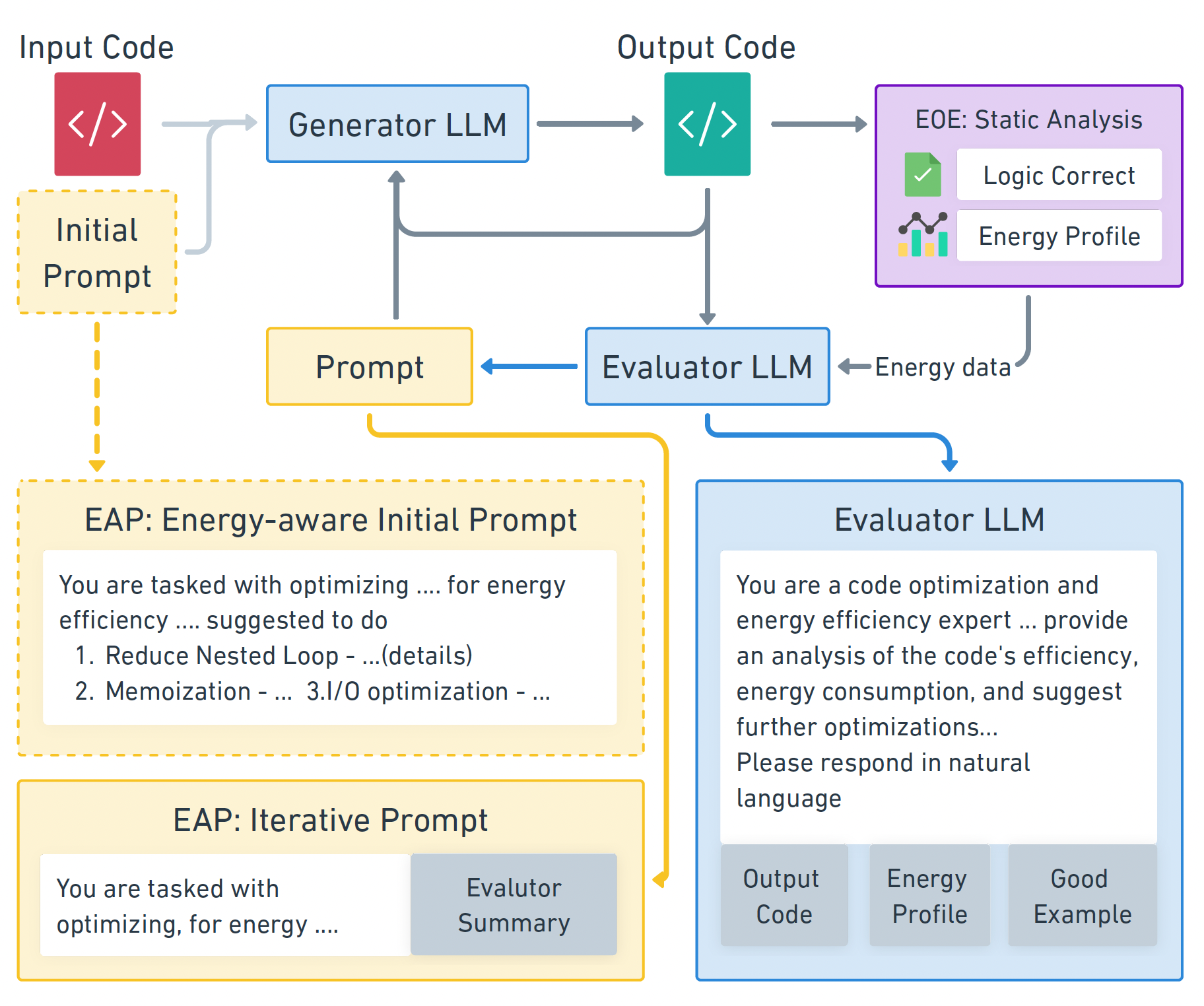}
    \centering
    \caption{
    Overview of our prototype.
      The core components of the automated feedback loop are
        Energy-Aware Prompting (EAP)
        and
        Energy Optimization Evaluation (EOE).
    }
    \label{fig:automated_LLM_tool}
\end{figure}

\cref{fig:automated_LLM_tool} shows the prototype design.
Following prior LLM applications, we chose an automated feedback loop structure to allow iterative improvement and repair.
Our loop involves two modules:
  Energy-Aware Prompting (EAP),
  and Energy Optimization Evaluation (EOE).
The EAP module integrates In-Context Learning~\cite{NEURIPS2020_1457c0d6} and Chain-of-Thought Prompting~\cite{Wei2024} to deliver detailed and informative instructions to the Generator LLM.
the EOE module evaluates the correctness and energy efficiency of the LLM-generated code, using an Evaluator LLM to provide feedback for iterative refinement.

\subsection{Prototype Implementation}


\bul{Automated Feedback Loop} 
This loop applies the GPT-4o as two agents, building on methods proposed in prior studies~\cite{qu2024recursive, madaan2023selfrefine, huang2023large}.
The AFL consists of a Generator LLM and an Evaluator LLM. As shown in \cref{fig:automated_LLM_tool}, the original code and Energy-Aware Prompting serve as inputs to the Generator LLM.
The Generator’s output is subsequently evaluated by the Evaluator LLM, which assesses the code based on its correctness and energy profile and provides recommendations for refinement.
By decoupling code generation from energy data analysis, we ensure that each LLM remains focused on its respective task, thereby improving overall performance. 

\bul{Energy-Aware Prompting (EAP)} 
In the Generator LLM, we employed One-Shot, Chain of Thought (CoT), and self-consistency prompting techniques to enhance the model's performance by providing sufficient contextual information on energy-efficient computing.
The prompt consists of three components.
First, the task for the Generator to perform is described.
Second, a brief code snippet, an optimized example, and a CoT rationale behind the optimization are provided.
Third, we integrate self-consistency by prompting the Generator to consider multiple optimization strategies focused on efficiency before selecting a final approach\cite{Wang2023}.
These prompts encourage the LLM to reason more carefully, and provide an efficient and correct solution.

\bul{Energy Optimization Evaluation (EOE)} 
This module has two objectives:
  (1) ensure the optimized code is compilable and functionally correct,
  and
  (2) measure energy efficiency and offer further optimization guidance.

The correctness evaluation involves two key steps.
First, the EOE module checks if the optimized code compiles successfully. If it fails, the Generator LLM receives the error message and optimized code for Self-Reflection and correction~\cite{chen2023teachinglargelanguagemodels, ma2024eureka}. Second, the EOE runs a regression test to verify logical correctness.
If runtime errors or output mismatches occur, the LLM is iteratively prompted with expected and actual outputs to refine optimization until the regression test passes, ensuring the optimized code matches the original functionality.

Once correctness is confirmed, we assess the energy efficiency and latency of the optimized code. The Evaluator LLM then analyzes the optimized code's energy profile and provides the Generator LLM with Natural Language Feedback (NLF) for further optimization.
The Evaluator is provided with (1) the original code, (2) the code with the lowest power consumption from the Optimization History Buffer, and (3) the most recent optimized code. This helps the Evaluator generate more precise feedback by correlating energy usage with optimizations while learning from the best historical example to drive further improvements.



{
\begin{table*}[t]
    \caption{
    Preliminary results of our prototype.
    We illustrate the performance of two optimization strategies (compiler, ours) across six algorithms compared with a baseline implementation.
    LLM improvements over the compiler optimizations are indicated with \better{green}, and degradation in \worse{red}.
    Values are rounded to the nearest integer.
    }
    \centering
    \arrayrulecolor[rgb]{0.502,0.502,0.502}
    \begin{tabular}{c|c|c|c|c|c|c|c}
        \toprule
        \small
        \multirow{2}{*}{\small \textbf{Program}} & \multirow{2}{*}{\textbf{Size}} & \multicolumn{2}{c|} {\textbf{Original Code}} & \multicolumn{2}{c|}{\textbf{Compiler Optimized}} & \multicolumn{2}{c}{\textbf{LLM Optimized}} \\ 
        & 
        & \textbf{Latency (ms)} & \textbf{Energy (J)} & \textbf{Latency (ms)} & \textbf{Energy (J)} & \textbf{Latency (ms)} & \textbf{Energy (J)} \\ 
        \toprule
        binary-trees & 139 lines (9 funcs) & 760 & 87 & 538 & 51 & \textcolor{ForestGreen}{187} -- \textcolor{Maroon}{954} & \textcolor{ForestGreen}{20} -- \textcolor{Maroon}{64} \\ 
        fannkuch-redux & 198 lines (5 funcs) & 6536 & 1119 & 1605 & 259 & \textcolor{Maroon}{5787} -- \textcolor{Maroon}{17918} & \textcolor{Maroon}{953} -- \textcolor{Maroon}{2945}\\ 
        n-body & 184 lines (6 funcs) & 20905 & 1150 & 2056 & 115 & \textcolor{Maroon}{11601} -- \textcolor{Maroon}{22575} & \textcolor{Maroon}{608} -- \textcolor{Maroon}{1193} \\ 
        pidigits & 68 lines (5 funcs) & 566 & 28 & 592 & 29 & \textcolor{ForestGreen}{525} -- \textcolor{ForestGreen}{542} & \textcolor{ForestGreen}{27} -- \textcolor{ForestGreen}{28} \\ 
        k-nucleotide & 154 lines (11 funcs) & 3685 & 476 & 865 & 85 & \textcolor{Maroon}{3677} -- \textcolor{Maroon}{5318} & \textcolor{Maroon}{478} -- \textcolor{Maroon}{789} \\ 
        spectral-norm & 135 lines (8 funcs) & 176 & 31 & 409 & 57 & \textcolor{ForestGreen}{87} -- \textcolor{Maroon}{6637} & \textcolor{ForestGreen}{15} -- \textcolor{Maroon}{342} \\ 
        \bottomrule
    \end{tabular}
    \label{table:table1}
\end{table*}
}

\subsection{Experimental Setup}
\bul{Software Benchmark}
Building on the prior research in Android energy efficiency~\cite{James2022, Oliveira2017}, we selected \textit{Energy-Language} as the benchmark tool for our experiments~\cite{Gouy_benchmarks_game}.
It offers comprehensive support for a wide range of testing algorithms across multiple programming languages and includes integrated energy measurement capabilities. 
Additionally, \textit{Energy-Language} is designed for scalability which facilitates easy incorporation of new codebases and programming languages into the testing suite. 

\bul{Power Consumption Measurement} 
We use the \textit{Energy-Language}'s built-in energy measurement system, which reads data from Model-Specific Registers (MSRs) supported by the Running Average Power Limit (RAPL) interface~\cite{thorat2017energy, hahnel2012measuring, hackenberg2015energy}.
To measure energy using MSRs, the RAPL registers accumulate energy usage over time. Energy efficiency is determined by reading the relevant MSR registers before and after a workload and calculating the difference.
These values are scaled to Joules using energy units from the \texttt{MSR\_RAPL\_POWER\_UNIT} register, allowing precise analysis of power usage for different components.

\subsection{Preliminary Results and Analysis}
We evaluated our prototype using the C++ benchmark, consisting of eleven unique programs. We successfully ran six of the programs with the results presented in \cref{table:table1}. 
The remaining five programs were not tested owing to compilation errors, runtime issues, or \code{GPT-4o} model token limit constraints. We collected data from three different optimization approaches: the original un-optimized program without the \texttt{-O3} flag, the compiler-optimized code with the \texttt{-O3} flag, and the LLM-optimized code without the \texttt{-O3} flag.


\bul{Comparison of Results with GCC --O3}
As shown in \cref{table:table1}, the results indicate that our framework reduces energy consumption in 83\% of the programs tested and outperforms the compiler optimization baseline in 50\% of them. 
Furthermore, the results from our framework are characterized by a much higher Standard Deviation relative to the compiler optimization baseline. This is most clearly seen in the \benchname{fannkuch-redux} test where the Standard Deviation for the measured energy and latency for our framework is 883 and 5441. It is also notable that, in the case of the \benchname{spectral-norm} benchmark, compiler optimizations increased energy consumption from 31J to 57J, while our code significantly reduced energy usage, bringing it down from 31J to just 15J.


\bul{Observations and Discussion}
Our prototype illustrates the potential for LLM-generated energy-aware optimizations.
We observed the LLM was proficient at improving memory traffic patterns, leading to reduced energy usage.
For instance, in \benchname{spectral-norm}, the LLM replaces Streaming SIMD Extensions (SSE) with Advanced Vector Extensions (AVX2), which performs twice as many floating-point computations in parallel. 
Additionally, the LLM-generated code processed data in larger chunks (size 256 bit vs. size 128 bit), reducing memory traffic and thus energy usage. These enhancements demonstrate the LLM's ability to optimize beyond the scope of individual functions. These results also highlight our framework's ability to reduce energy usage in a deliberate manner.

We highlight two limitations of the prototype.
First, there is significant variance in the energy consumption of LLM-optimized software across successive iterations.
We conjecture two causes of this property.
First, LLM output is sensitive to hyperparameter choices such as Top-K and Temperature; therefore the input-output relationship can be non-deterministic.
Though this non-determinism can be controlled (\eg with Temperature$=0$) using a non-zero Temperature allows for more substantial optimization attempts.
Second, we hypothesize that prompting alone is insufficient.
Our prototype relies on prompting and learning from feedback, but uses existing model parameters.
As a result, the LLM may not be properly conditioned on applying energy-aware optimizations.

As a second limitation, the Generator LLM was effective at resolving compiler errors, but struggled to address run-time errors. 
Unlike compiler errors, run-time errors often do not yield an explicit error message. 
As a result, in longer and more complex programs such as \benchname{fannkuch-redux}, \benchname{n-body}, and \benchname{k-nucleotide}, the LLM required more iterations to address this type of error. 
Additionally, we observed that spending more than three iterations on code-refinement led an increase in energy usage and latency.
This indicates that the LLM lacks an understanding of run-time error-handling or early-stopping criteria for code-optimization.

\section{Future Plans}\label{sec:plan}
\subsection{Improvements to Evaluation}

\bul{Baseline Analysis}
In our preliminary evaluation, we used the \code{gcc -O3} optimization flag for C++ benchmarks. We will extend similar analysis to other languages, such as Java with its \code{JIT} optimizer. Additionally, we will benchmark our approach against the recently released \code{GPT-o1} model which has native chain-of-thought ability. 
We will evaluate our pipeline on larger applications including standard data center benchmarks such as the Facebook benchmark of datacenter cloud applications, 
\textit{DCPerf}~\cite{DcPerf2023}, and standard HPC benchmarks evaluated in the Green500 work~\cite{Feng2007}.

\bul{Extended SOTA Comparison}
We plan to evaluate state-of-the-art code-generation models on baseline programs, and compare their performance with our prototype. 
This will identify shortcomings and potential improvements for energy-aware LLM code generation, paving the way for additional research efforts.
It will also be interesting to compare with other approaches for energy efficiency, such as search-based software engineering approaches.

\subsection{Prototype Framework Refinements} 
\bul{LLM Prompting} We will refine our prompts.
Building on previous work, we will apply prompt engineering methods such as few-shot learning, structured CoT prompting~\cite{Li2023}, and prompt chaining.
This will yield the first catalog of LLM prompts for energy efficiency.

\bul{LLM Fine-tuning}
We will fine-tune the LLM on a curated dataset of energy-efficient algorithms and low-power software practices. 
For this purpose, we will develop a new training dataset for energy-efficient computing, focusing on efficiency-critical data center software and HPC algorithms.
Inspired by prior work~\cite{moura2015miningEnergyAwareCommits}, we will perform large-scale data collection from GitHub repositories, using a language model to analyze commit messages and identify energy-efficiency commits.
This dataset would allow us to fine-tune the LLM performance.


\bul{Hardware-Aware Code Generation}
Our current approach focuses on optimizing specific components of the system or hardware at the code level without considering its deployment scenario (system resources, hardware architecture, etc.).
We believe integrating a Retrieval-Augmented-Generation (RAG) framework~\cite{lewis2020RAG} that relates code-modifications to hardware and system specifications could rectify this issue.
In \benchname{spectral-norm}, we observed GPT do this without specific prompting.

\subsection{Broader Research Directions}
\bul{Weighing Costs of LLM-Driven Energy Optimizations}
AI systems such as LLMs are power hungry.
We will assess both the energy and computational costs of using LLMs for software energy-efficiency optimization, comparing the energy usage of different LLMs for the same tasks. 
Based on the findings, we will develop adaptive strategies, such as triggering LLM-based optimizations only when significant energy savings are expected.

\bul{Multi-Objective Optimization}
We will explore whether LLMs can be adapted for multi-objective optimization, balancing energy efficiency with key performance metrics like latency and throughput. Developing tools that optimize across these competing objectives would enhance LLM applicability in real-world scenarios where such trade-offs are critical.
LLMs could also generate code across a spectrum of efficiency constraints.
In this context, formal verification techniques might be incorporated for stronger correctness guarantees~\cite{tihanyi2023formai}.


\bul{Second-Class Citizenship is Better than None}
After decades of calls for energy-aware computing, it seems clear to us that energy will not be prioritized as much as business-critical metrics such as latency and throughput.
We hope, however, that it will not be ignored.
We suggest that engineers and researchers might approach energy as a secondary performance metric, one that should still be considered after primary metrics are met.
This would require engineering tools and processes that support energy considerations in a lightweight way, and that can be applied after the primary engineering goals are met.
Our LLM approach may become one such tool.


\section{Conclusion}
In this paper, we explored the potential of applying LLMs to energy-efficient software development and introduced an automated approach for energy-aware code optimization.
Our initial experiments demonstrate that LLMs can improve energy efficiency while maintaining code correctness.
Based on this emerging result, we discuss our immediate and longer-term research plans.

\vspace{0.2cm}
\uline{Data availability:} Code, prompts, and data are available: \url{https://anonymous.4open.science/r/E2COOL-5CD4}.

\clearpage
\balance

\bibliographystyle{IEEEtran}
\bibliography{refs/references}

\end{document}